\documentclass[twocolumn]{aastex631}

\shortauthors{IceCube Collaboration}

\begin{document}

\title{Search for 10--1000 GeV neutrinos from Gamma Ray Bursts with IceCube}

\affiliation{III. Physikalisches Institut, RWTH Aachen University, D-52056 Aachen, Germany}
\affiliation{Department of Physics, University of Adelaide, Adelaide, 5005, Australia}
\affiliation{Dept. of Physics and Astronomy, University of Alaska Anchorage, 3211 Providence Dr., Anchorage, AK 99508, USA}
\affiliation{Dept. of Physics, University of Texas at Arlington, 502 Yates St., Science Hall Rm 108, Box 19059, Arlington, TX 76019, USA}
\affiliation{CTSPS, Clark-Atlanta University, Atlanta, GA 30314, USA}
\affiliation{School of Physics and Center for Relativistic Astrophysics, Georgia Institute of Technology, Atlanta, GA 30332, USA}
\affiliation{Dept. of Physics, Southern University, Baton Rouge, LA 70813, USA}
\affiliation{Dept. of Physics, University of California, Berkeley, CA 94720, USA}
\affiliation{Lawrence Berkeley National Laboratory, Berkeley, CA 94720, USA}
\affiliation{Institut f{\"u}r Physik, Humboldt-Universit{\"a}t zu Berlin, D-12489 Berlin, Germany}
\affiliation{Fakult{\"a}t f{\"u}r Physik {\&} Astronomie, Ruhr-Universit{\"a}t Bochum, D-44780 Bochum, Germany}
\affiliation{Universit{\'e} Libre de Bruxelles, Science Faculty CP230, B-1050 Brussels, Belgium}
\affiliation{Vrije Universiteit Brussel (VUB), Dienst ELEM, B-1050 Brussels, Belgium}
\affiliation{Department of Physics and Laboratory for Particle Physics and Cosmology, Harvard University, Cambridge, MA 02138, USA}
\affiliation{Dept. of Physics, Massachusetts Institute of Technology, Cambridge, MA 02139, USA}
\affiliation{Dept. of Physics and The International Center for Hadron Astrophysics, Chiba University, Chiba 263-8522, Japan}
\affiliation{Department of Physics, Loyola University Chicago, Chicago, IL 60660, USA}
\affiliation{Dept. of Physics and Astronomy, University of Canterbury, Private Bag 4800, Christchurch, New Zealand}
\affiliation{Dept. of Physics, University of Maryland, College Park, MD 20742, USA}
\affiliation{Dept. of Astronomy, Ohio State University, Columbus, OH 43210, USA}
\affiliation{Dept. of Physics and Center for Cosmology and Astro-Particle Physics, Ohio State University, Columbus, OH 43210, USA}
\affiliation{Niels Bohr Institute, University of Copenhagen, DK-2100 Copenhagen, Denmark}
\affiliation{Dept. of Physics, TU Dortmund University, D-44221 Dortmund, Germany}
\affiliation{Dept. of Physics and Astronomy, Michigan State University, East Lansing, MI 48824, USA}
\affiliation{Dept. of Physics, University of Alberta, Edmonton, Alberta, T6G 2E1, Canada}
\affiliation{Erlangen Centre for Astroparticle Physics, Friedrich-Alexander-Universit{\"a}t Erlangen-N{\"u}rnberg, D-91058 Erlangen, Germany}
\affiliation{Physik-department, Technische Universit{\"a}t M{\"u}nchen, D-85748 Garching, Germany}
\affiliation{D{\'e}partement de physique nucl{\'e}aire et corpusculaire, Universit{\'e} de Gen{\`e}ve, CH-1211 Gen{\`e}ve, Switzerland}
\affiliation{Dept. of Physics and Astronomy, University of Gent, B-9000 Gent, Belgium}
\affiliation{Dept. of Physics and Astronomy, University of California, Irvine, CA 92697, USA}
\affiliation{Karlsruhe Institute of Technology, Institute for Astroparticle Physics, D-76021 Karlsruhe, Germany}
\affiliation{Karlsruhe Institute of Technology, Institute of Experimental Particle Physics, D-76021 Karlsruhe, Germany}
\affiliation{Dept. of Physics, Engineering Physics, and Astronomy, Queen's University, Kingston, ON K7L 3N6, Canada}
\affiliation{Department of Physics {\&} Astronomy, University of Nevada, Las Vegas, NV 89154, USA}
\affiliation{Nevada Center for Astrophysics, University of Nevada, Las Vegas, NV 89154, USA}
\affiliation{Dept. of Physics and Astronomy, University of Kansas, Lawrence, KS 66045, USA}
\affiliation{Centre for Cosmology, Particle Physics and Phenomenology - CP3, Universit{\'e} catholique de Louvain, Louvain-la-Neuve, Belgium}
\affiliation{Department of Physics, Mercer University, Macon, GA 31207-0001, USA}
\affiliation{Dept. of Astronomy, University of Wisconsin{\textemdash}Madison, Madison, WI 53706, USA}
\affiliation{Dept. of Physics and Wisconsin IceCube Particle Astrophysics Center, University of Wisconsin{\textemdash}Madison, Madison, WI 53706, USA}
\affiliation{Institute of Physics, University of Mainz, Staudinger Weg 7, D-55099 Mainz, Germany}
\affiliation{Department of Physics, Marquette University, Milwaukee, WI 53201, USA}
\affiliation{Institut f{\"u}r Kernphysik, Westf{\"a}lische Wilhelms-Universit{\"a}t M{\"u}nster, D-48149 M{\"u}nster, Germany}
\affiliation{Bartol Research Institute and Dept. of Physics and Astronomy, University of Delaware, Newark, DE 19716, USA}
\affiliation{Dept. of Physics, Yale University, New Haven, CT 06520, USA}
\affiliation{Columbia Astrophysics and Nevis Laboratories, Columbia University, New York, NY 10027, USA}
\affiliation{Dept. of Physics, University of Oxford, Parks Road, Oxford OX1 3PU, United Kingdom}
\affiliation{Dipartimento di Fisica e Astronomia Galileo Galilei, Universit{\`a} Degli Studi di Padova, I-35122 Padova PD, Italy}
\affiliation{Dept. of Physics, Drexel University, 3141 Chestnut Street, Philadelphia, PA 19104, USA}
\affiliation{Physics Department, South Dakota School of Mines and Technology, Rapid City, SD 57701, USA}
\affiliation{Dept. of Physics, University of Wisconsin, River Falls, WI 54022, USA}
\affiliation{Dept. of Physics and Astronomy, University of Rochester, Rochester, NY 14627, USA}
\affiliation{Department of Physics and Astronomy, University of Utah, Salt Lake City, UT 84112, USA}
\affiliation{Dept. of Physics, Chung-Ang University, Seoul 06974, Korea}
\affiliation{Oskar Klein Centre and Dept. of Physics, Stockholm University, SE-10691 Stockholm, Sweden}
\affiliation{Dept. of Physics and Astronomy, Stony Brook University, Stony Brook, NY 11794-3800, USA}
\affiliation{Dept. of Physics, Sungkyunkwan University, Suwon 16419, Republic of Korea}
\affiliation{Institute of Physics, Academia Sinica, Taipei, 11529, Taiwan}
\affiliation{Dept. of Physics and Astronomy, University of Alabama, Tuscaloosa, AL 35487, USA}
\affiliation{Dept. of Astronomy and Astrophysics, Pennsylvania State University, University Park, PA 16802, USA}
\affiliation{Dept. of Physics, Pennsylvania State University, University Park, PA 16802, USA}
\affiliation{Dept. of Physics and Astronomy, Uppsala University, Box 516, SE-75120 Uppsala, Sweden}
\affiliation{Dept. of Physics, University of Wuppertal, D-42119 Wuppertal, Germany}
\affiliation{Deutsches Elektronen-Synchrotron DESY, Platanenallee 6, D-15738 Zeuthen, Germany}

\author[0000-0001-6141-4205]{R. Abbasi}
\affiliation{Department of Physics, Loyola University Chicago, Chicago, IL 60660, USA}

\author[0000-0001-8952-588X]{M. Ackermann}
\affiliation{Deutsches Elektronen-Synchrotron DESY, Platanenallee 6, D-15738 Zeuthen, Germany}

\author{J. Adams}
\affiliation{Dept. of Physics and Astronomy, University of Canterbury, Private Bag 4800, Christchurch, New Zealand}

\author[0000-0002-9714-8866]{S. K. Agarwalla}
\altaffiliation{also at Institute of Physics, Sachivalaya Marg, Sainik School Post, Bhubaneswar 751005, India}
\affiliation{Dept. of Physics and Wisconsin IceCube Particle Astrophysics Center, University of Wisconsin{\textemdash}Madison, Madison, WI 53706, USA}

\author[0000-0003-2252-9514]{J. A. Aguilar}
\affiliation{Universit{\'e} Libre de Bruxelles, Science Faculty CP230, B-1050 Brussels, Belgium}

\author[0000-0003-0709-5631]{M. Ahlers}
\affiliation{Niels Bohr Institute, University of Copenhagen, DK-2100 Copenhagen, Denmark}

\author[0000-0002-9534-9189]{J.M. Alameddine}
\affiliation{Dept. of Physics, TU Dortmund University, D-44221 Dortmund, Germany}

\author{N. M. Amin}
\affiliation{Bartol Research Institute and Dept. of Physics and Astronomy, University of Delaware, Newark, DE 19716, USA}

\author[0000-0001-9394-0007]{K. Andeen}
\affiliation{Department of Physics, Marquette University, Milwaukee, WI 53201, USA}

\author[0000-0003-2039-4724]{G. Anton}
\affiliation{Erlangen Centre for Astroparticle Physics, Friedrich-Alexander-Universit{\"a}t Erlangen-N{\"u}rnberg, D-91058 Erlangen, Germany}

% ...

\collaboration{408}{IceCube Collaboration}

\begin{abstract}
We present the results of a search for 10--1000~GeV neutrinos from 2268 gamma-ray bursts over 8 years of IceCube-DeepCore data. This work probes burst physics below the photosphere where electromagnetic radiation cannot escape. Neutrinos of tens of GeVs are predicted in sub-photospheric collision of free streaming neutrons with bulk-jet protons. In a first analysis, we searched for the most significant neutrino-GRB coincidence using six overlapping time windows centered on the prompt phase of each GRB. In a second analysis, we conducted a search for a group of GRBs, each individually too weak to be detectable, but potentially significant when combined. No evidence of neutrino emission is found for either analysis. The most significant neutrino coincidence is for \textit{Fermi-GBM} GRB bn~140807500, with a $p$-value of 0.097 corrected for all trials. The binomial test used to search for a group of GRBs had a $p$-value of 0.65 after all trial corrections. The binomial test found a group consisting only of GRB bn~140807500 and no additional GRBs. The neutrino limits of this work complement those obtained by IceCube at TeV to PeV energies. We compare our findings for the large set of GRBs as well as GRB~221009A to the sub-photospheric neutron-proton collision model and find that GRB~221009A provides the most constraining limit on baryon loading. For a jet Lorentz factor of 300 (800), the baryon loading on GRB~221009A is lower than 3.85 (2.13) at a 90\% confidence level.
\end{abstract}


\section{Introduction} \label{sec:intro}

Gamma Ray Bursts, or GRBs, \citep{2004RvMP...76.1143P} are among the most powerful objects in the Universe. During the milliseconds to hundreds of seconds ``prompt phase'', copious amounts of keV--MeV photons are released. The fireball scenario \citep{1999PhR...314..575P} assumes a radiation-dominated electron-positron plasma ejected in a jet at a high Lorentz factor, $\Gamma\gtrsim 300$. The prompt phase is followed by a decaying multi-wavelength afterglow that can be observed from radio to X-rays for days to years \citep{2021ApJ...911...14K}. GRBs are empirically classified as either long for a duration of the prompt phase $T_{90}$, larger than 2 seconds, or as short otherwise. Long GRBs are associated with supernovae, see e.g. \citet{2013RSPTA.37120275H} and \citet{2017AdAst2017E...5C} for reviews. Short GRBs are associated with the merger of compact objects as corroborated with the observation of gravitational waves for GRB/GW~170817A \citep{PhysRevLett.119.161101}. 

GRBs have been proposed as cosmic ray accelerators \citep{1995ApJ...453..883V, 1995PhRvL..75..386W}. The interaction of these cosmic rays with local environment radiation and/or matter would result in PeV neutrinos during the prompt phase \citep{PhysRevLett.78.2292,2013PhRvL.110l1101Z}, TeV precursor neutrinos prior to the prompt phase \citep{2001PhRvL..87q1102M,2003PhRvD..68h3001R,2013PhRvL.111l1102M} and EeV neutrinos during the afterglow \citep{2000ApJ...541..707W}. IceCube has discovered an all-sky, all-flavor, flux of neutrinos from $\sim$10~TeV to $\sim$10~PeV \citep{PhysRevLett.125.121104,2021PhRvD.104b2002A,2022ApJ...928...50A}. GRBs are a long-standing candidate to explain, at least in part, this flux.

Neutrons and protons entrained on a GRB fireball can lead to GeV-scale neutrinos \citep{PhysRevLett.85.1362}. Sub-photospheric neutron-proton decoupling in the fireball leads to 1--10 GeV neutrinos. These neutrinos are too low energy for the study reported here. Decoupled neutrons collide below the photosphere with fireball-entrained protons \citep{2013PhRvL.111m1102M}. These collisions produce charged and neutral pions. Charged pions decay, directly and indirectly, into neutrinos and charged leptons. In the collision scenario, typical neutrino energy is tens of GeV, an energy range that is tested in this work. For both decoupling and collision scenarios, neutrinos have a quasi-thermal spectrum and precede the prompt phase by tens of seconds. 

To date, there is no evidence of neutrino emission from GRBs. IceCube has conducted searches for TeV to PeV neutrinos in coincidence with GRBs during the prompt phase \citep{Nature-GRB-IceCube,2015ApJ...805L...5A,2016ApJ...824..115A,2017ApJ...843..112A}. Recently, IceCube has searched for neutrino-GRB coincidences in the TeV to PeV energy range during the prompt, precursor, and afterglow phases \citep{2022ApJ...939..116A}. 
During the prompt phase, IceCube results demonstrated that GRBs cannot be responsible for more than $\sim$1\% of IceCube's extragalactic neutrino flux. For time windows around GRBs of $10^4$~s, centered on the prompt phase, GRBs cannot contribute to more than 24\% of the extragalactic neutrino flux.
ANTARES, which operated in the TeV to PeV energy range, conducted searches for neutrinos in coincidence with GRBs without finding a coincidence \citep{10.1093/mnras/stx902,2021JCAP...03..092A,2021MNRAS.500.5614A}. Super-Kamiokande has conducted a search for neutrinos from GRBs above 8~MeV also with null results \citep{2021arXiv210103480T}. 

The brightest GRB of all time, GRB 221009A \citep{2023ApJ...946L..31B}, was detected while this publication was being prepared and is not included in the list of 2268 GRBs we analyze here. \citet{2022ApJ...941L..10M} have calculated neutrino fluxes for GRB~221009A in the sub-photospheric neutron-proton collision model.
GRB~221009A was studied in neutrinos from MeV to PeV by IceCube \citep{2023ApJ...946L..26A,icrc2023:grb221009a} and no evidence for neutrino emission was found. The search for 10--1000 GeV neutrinos from GRB 221009A by IceCube is methodologically identical to the work presented here, differing only on the duration of time windows used. GRB~221009A was also studied by KM3NeT and no evidence for neutrino emission was found either \citep{2022GCN.32741....1K}.

In this work, we present a study of 2268 GRBs detected by satellite-borne instruments with eight years of IceCube-DeepCore data in the 10~GeV to 1000~GeV energy range. The present study covers the precursor, prompt, and early afterglow phases of GRBs for a total duration of up to 500~s centered on the prompt phase. We find no evidence for 10--1000 GeV neutrino emission by GRBs. 
We use two analysis methods, one to search for the most significant GRB-neutrino coincidence and another to search for an ensemble of GRBs that may be too weak to be detectable individually with neutrinos but may be significant as a group.
We find no evidence of neutrino emission for these 2268 GRBs in either analysis and we set limits on the time-integrated, all neutrino flavor, neutrino flux. The results presented here are compared to limits for $\gtrsim$TeV that IceCube has published. 
%
%
Because GRB~221009A has an extremely high energy fluence, we compare predicted signal expectations for the sub-photospheric neutron-proton collision model for 2264 GRBs, for which energy fluence has been reported, to GRB~221009A. 
We find that, under the model assumptions, the signal expectation for GRB~221009A is a factor 6--8 higher than for the other 2264 GRBs combined. Also, the total background for GRB~221009A is significantly lower. Thus, we derive the best possible limit on jet baryon loading using neutrino limits on GRB~221009A by IceCube \citep{icrc2023:grb221009a}. 
Assuming a jet Lorentz factor of 300 (800), the baryon loading on GRB 221009A is lower than 3.85 (2.13) at a 90\% confidence level.

\section{IceCube, DeepCore, and Dataset description} 

The IceCube Neutrino Observatory \citep{2017JInst..12P3012A} consists of an array of 5,160 digital optical modules (DOMs) on a total of 86 strings embedded within one cubic kilometer of Antarctic ice at the South Pole. All DOMs include a downward-facing photomultiplier tube (PMT) \citep{2020JInst..15P6032A}, and associated electronics \citep{2009NIMPA.601..294A} enclosed within a glass vessel and are deployed from 1.45~km to 2.45~km below the surface. IceCube is optimized for $\gtrsim$TeV observations, matching an inter-string spacing of $\sim$125 m. 
Six \textit{DeepCore strings} were installed on the vertices of a hexagon with a side of 42~m. At the hexagon center is the central standard IceCube string. Two additional \textit{DeepCore-infill strings} have been placed inside the hexagon with even smaller horizontal separation.
The combination of these 8 strings and 7 nearby standard IceCube strings form the 
DeepCore sub-detector \citep{2012APh....35..615A}. 
The physics region of DeepCore is a cylinder of 125~m in radius and 350~m in height. DeepCore is optimized for $\gtrsim$10~GeV neutrino observations.
For DeepCore and DeepCore-infill strings, 50 DOMs are installed with 7~m vertical spacing, between 2.1~km and 2.45~km of depth. The other 10 DOMs are installed with a 10~m vertical separation between 1.8~km and 1.9~km of depth. These DOMs are used for enhanced down-going cosmic ray muon rejection. The region between 2,000~m and 2,100~m is not instrumented in DeepCore and DeepCore-infill strings as glacial ice in this region has worse optical properties \citep{2013NIMPA.711...73A}. All the DOMs in the DeepCore strings are equipped with high quantum efficiency PMTs. The DOMs on the DeepCore-infill strings have a mixture of standard and high quantum efficiency PMTs.

We use IceCube-DeepCore data collected between April 26, 2012, and May 29, 2020, with a total live time of 7.68 years, which corresponds to an uptime of 95\%. The data sample used in this publication, called GRECO-Astronomy (GeV Reconstructed Events with Containment for Oscillations), is described in detail in \citet{2022arXiv221206810A}. GRECO-Astronomy has sensitivity over the entire sky (4$\pi$~sr) that is only weakly dependent on declination. As they produce virtually identical signatures, IceCube-DeepCore cannot distinguish $\nu$ from $\bar{\nu}$. The GRECO-Astronomy dataset is sensitive to all neutrino flavors via cascades and starting tracks. 

\begin{figure*}[t]
\centering
\includegraphics[width=0.49\textwidth]{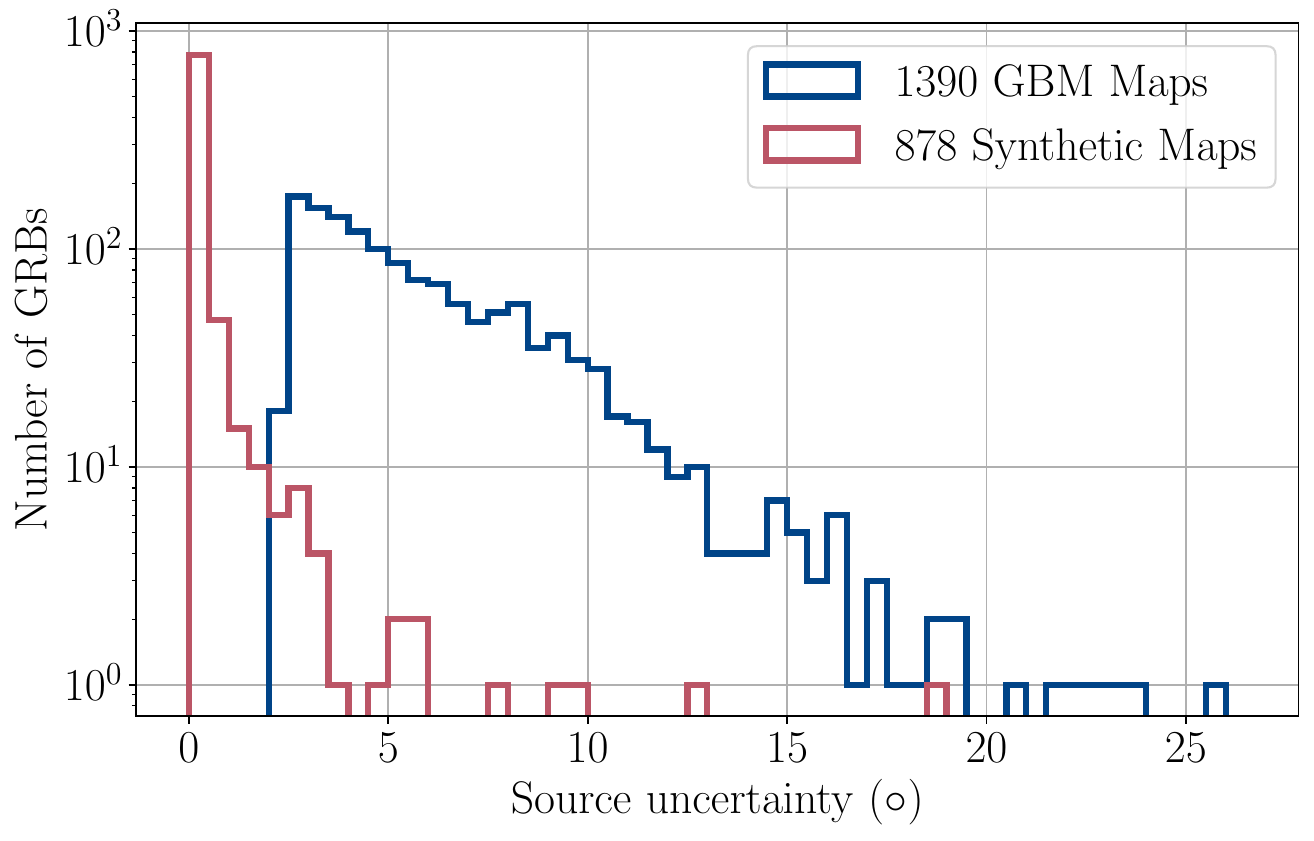}
\includegraphics[width=0.49\textwidth]{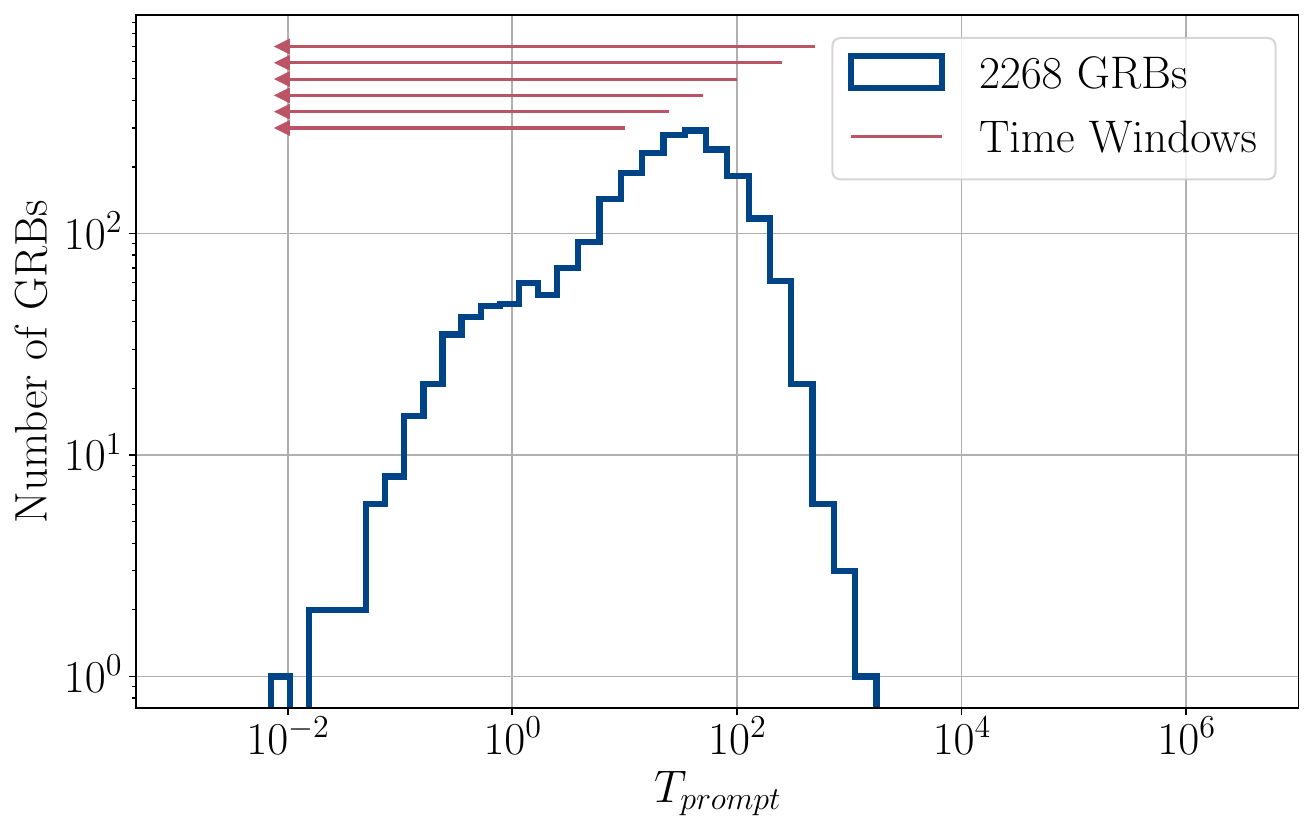}
\caption{(Left) Distribution of circularized 1-sigma localization uncertainties for GRBs used in this work. (Right) Distribution of prompt phase duration, $T_{prompt}$, for GRBs in this study. See text for the definition of $T_{prompt}$. Also shown, as arrows, are the durations of the six time windows used to search for neutrino-GRB coincidences. It can be seen that the widest time window, of duration 500~s, covers the $T_{prompt}$ of all GRBs but ten. \label{fig:grb-durations-uncertainty}}
\end{figure*}

Cascades, or showers, are the product of neutral current interactions for all neutrino flavors, as well as charged current interactions of $\nu_e$ and the majority of $\nu_\tau$. For cascades, all energy is deposited in a small volume that is contained in DeepCore. This leads to good energy resolution, but poor angular resolution \citep{2022arXiv221206810A}. 
Starting tracks are predominantly muons originating from $\nu_\mu$ charged-current interactions. In this case, part of the energy is deposited in a shower at the neutrino interaction vertex, plus an outgoing, relatively long-range muon. If sufficiently long, tracks can provide smaller angular uncertainty estimates than cascades. The GRECO-Astronomy dataset includes event reconstructions for both cascades and starting tracks. The median angular resolution ranges from $\sim$40~degrees for cascades with reconstructed visible energy of 10~GeV to a few degrees for starting tracks with $\gtrsim$100~GeV reconstructed energy. In this work, we keep only events with reconstructed energy above 10~GeV, which reduces the average GRECO-Astronomy event rate from 4.6~mHz (16.6 per hour) to 4.17~mHz (15 per hour). This choice is taken because the angular uncertainty of lower reconstructed energy events is judged too poor to be used here. There are 1,010,151 events in the final data sample used in this work.

We used GRBweb \citep{grbweb} to select GRBs for this study. This catalog brings together information from a variety of public GRB databases, such as the General Coordinate Network (GCN) Notices and Circulars, and compiles electromagnetic observational data from a large number of instruments including \textit{Fermi-GBM} \citep{2009ApJ...702..791M}, \textit{Fermi-LAT} \citep{2009ApJ...697.1071A}, \textit{Swift-BAT} \citep{2005SSRv..120..143B}, the \textit{IPN network} \citep{2013ApJS..207...39H}, etc. Within the time period studied, 2297 GRBs were recorded in GRBweb. Of these, we have selected 2268 GRBs for this work. We have excluded 29 GRBs that do not have a sky localization or localization uncertainty. For each GRB we obtain four pieces of information from GRBweb: the sky localization, the localization uncertainty, the duration, and the energy fluence. When multiple detectors observe a given burst, we use the localization information for the detector with the smallest localization uncertainty. In GRBweb, localization uncertainty is assumed to be the 1$\sigma$ uncertainty for a 2D normal distribution, implying that the true position of the GRB lies in the uncertainty circle 39\% of the time. Some instruments, e.g., \textit{Fermi-GBM} provide more detailed information than the 1$\sigma$ localization uncertainty. For these cases, GRBweb calculates the equivalent 1$\sigma$ circularized uncertainty that provides the correct coverage. For GRBs localized by \textit{IPN}, it is not possible to calculate 1$\sigma$ circularized uncertainty since these GRBs are located in a box in the sky and only 3$\sigma$ bounds are provided.
For \textit{IPN}, we use the larger of the two uncertainty box dimensions to generate a circularized uncertainty region.
Each instrument that observes a GRB will typically provide a start time, $T_0$, and duration of the prompt phase, $T_{90}$. GRBweb determines the largest duration that covers all the $T_{90}$ observations. We call this duration $T_{prompt}$. 
In the majority of bursts used in this work $T_{prompt}$ matches the $T_{90}$ of a single observing instrument, which is most frequently \textit{Fermi-GBM}. The energy fluence $f_\gamma$, during the prompt phase, is chosen by GRBweb to be that of the instrument with the widest energy observation band. Often this will be \textit{Fermi-GBM}, as it has sensitivity from 10~keV to 10~MeV. The energy fluence is not used in the search for neutrino-GRB coincidences, but it is used in the interpretation of the results.

Each burst location and localization uncertainty is considered as a prior described by a probability density function. When the best localization of a given burst is by \textit{Fermi-GBM}, we use HEALPix maps \citep{2005ApJ...622..759G} produced by \textit{Fermi-GBM}. \textit{Fermi-GBM} maps starting 
in early 2018 were released publicly \citep{2020ApJ...895...40G}. Maps for bursts prior to 2018 were processed similarly using the same GBM Data Tools. However, the metadata in the files have not been fully qualified and the files have not been uploaded to HEASARC \footnote{FTP data: \url{https://heasarc.gsfc.nasa.gov/FTP/fermi/data/gbm/triggers/}}. Therefore, we use preliminary maps by \citet{GoldsteinWood}. If a different instrument provides a better localization than \textit{Fermi-GBM}, then synthetic HEALpix maps were created. These synthetic maps have uniform (top-hat) probability over the circularized 1$\sigma$ localization uncertainty obtained via GRBweb. If the GRBweb localization uncertainty is smaller than 1 degree, then the synthetic map is created with a radius of 1 degree. As can be seen on the left panel of Figure \ref{fig:grb-durations-uncertainty}, this lower bound of 1 degree is used in the vast majority of synthetic maps. The lower bound of 1 degree is used for computational reasons and does not significantly affect this analysis, as the typical GRECO-Astronomy event has an angular uncertainty much larger than 1 degree. 

Of the 2268 GRBs studied in this work, 1,390 have \textit{Fermi-GBM} HEALPix skymaps and 878 have synthetic skymaps. All HEALPix maps have been rebinned to a setting of Nside=64, corresponding to 49,152 pixels over the entire sky. Each pixel has an angular area of $\sim$0.84 square degrees, which is usually smaller than the angular resolution for the best-reconstructed events in the GRECO-Astronomy dataset. 

\section{Methods} \label{sec:methods}

We have conducted two analyses in this work.
First, we search for the statistically most significant GRB in a temporal and directional coincidence with GRECO-Astronomy neutrino events. Second, we search for the most statistically significant group of GRBs in temporal and directional correlation with GRECO-Astronomy neutrinos. This latter search can potentially find a subset of the 2268 GRBs as neutrino emitters, even though none of the GRBs were statistically significant on their own. This second analysis uses a binomial test to statistically combine the results of the first analysis.    

\subsection{Search for the most significant neutrino-GRB correlation}

Each GRB is studied using six pre-defined time windows with durations of 10~s, 25~s, 50~s, 100~s, 250~s, and 500~s centered in the middle of the $T_{prompt}$ window.  These overlapping time windows cover or over-cover the vast majority of prompt phases of the 2268 GRBs in our list. Longer-than-prompt time windows are used to identify potential precursor and early afterglow neutrino-GRB correlations. The right panel of Figure \ref{fig:grb-durations-uncertainty} shows the distribution of $T_{prompt}$ for all the GRBs used in this work, as well as the duration of the time windows that we study. To characterize GRB neutrino emission, we use the time-integrated neutrino number flux:
\begin{equation}
\label{eq:numberflux}
F(E) = \int^{t_b}_{t_a} \frac{dN_{\nu+\bar{\nu}}}{dEdAdt} dt,
\end{equation}
where $t_a$ and $t_b$ set the time window being considered and $dN_{\nu+\bar{\nu}}/dEdAdt$ is the number of neutrinos per unit of area, energy, and time. Equal values of $F(E)$ can be obtained for arbitrary neutrino time profiles within $t_a$ and $t_b$.

The unbinned maximum likelihood method is a common approach to search for time-dependent neutrino sources \citep{BRAUN2008299}. In this work, we use an extended likelihood, which is a function of the number of neutrino signal events, $n_s$, the location of the sky that is being studied, $\vec{\Omega}$, conditional on the expected number of background events, $n_b$, and per-neutrino information  $\mathbf{x}_i$. The neutrino information $\mathbf{x}_i$ is reconstructed direction (right ascension, $\alpha_i$ and declination, $\delta_i$), neutrino angular uncertainty estimator $\sigma_i$, neutrino arrival time $t_i$, and reconstructed energy $E_i$. The extended likelihood function is:
\begin{eqnarray}
&& \mathcal{L} (\vec{\Omega},n_s | n_b, \left \{ \mathbf{x}_i \right \}) = 
 \frac{(n_s+n_b)^N e^{-(n_s+n_b)}}{N!} \nonumber \\
&& \times \prod_{i=1}^N \left(  \frac{n_s }{n_s+n_b}\mathcal{S} (\mathbf{x}_i,\vec{\Omega})+ \frac{n_b }{n_s+n_b}\mathcal{B} (\mathbf{x}_i,\vec{\Omega})  \right) , 
\end{eqnarray}
with the index $i$ iterating over all $N$ candidate neutrino events in a given time window.
The Poisson term is frequently used by IceCube in transient neutrino analyses, including GRB studies \citep{2015ApJ...805L...5A,2016ApJ...824..115A,2017ApJ...843..112A,2022ApJ...939..116A}, in which $n_s+n_b$ is a relatively small number. $\mathcal{S}(\cdot)$ and $\mathcal{B}(\cdot)$ are the signal and background probability density functions (PDFs) respectively. Both $\mathcal{S}(\cdot)$ and $\mathcal{B}(\cdot)$ are described as the product of a directional term and an energy term.

The directional term of $\mathcal{S}(\cdot)$ is represented by a Kent function \citep{kent1982fisher}. 
\begin{equation}
S_{space} (\Delta\psi_i) = \frac{\kappa}{4\pi \sinh \kappa} e^{ \kappa \cos \Delta\psi_i}, 
\end{equation}
where $\Delta\psi_i$ is the angular difference between each reconstructed event direction and the direction, $\vec{\Omega}$, being studied and $\kappa = 1/\sigma_{i}^2$. The choice of a Kent function, instead of a normal distribution is made because the angular uncertainty of GRECO-Astronomy events is relatively large. The background space PDF, $B_{i,space}$, is a function of zenith only due to the approximate azimuthal symmetry of the IceCube-DeepCore detector, and is determined from 7.68~years of data, but excluding events from the corresponding time window around all of the GRBs. Background events in GRECO-Astronomy are $\sim$60\% due to atmospheric neutrinos, and $\sim$40\% due to down-going atmospheric muons \citep{2022arXiv221206810A}.  

The energy term of the signal PDF is determined from simulations that assume an energy spectrum of $E^{-2.5}$. This choice of spectrum was made so that the most sensitive energy range of the study would correspond to approximately 20~GeV to 500~GeV. The energy term of the background PDF is determined from 7.68~years of data, again, excluding events from the time window around all of the GRBs. 

For each of the six time windows of a GRB, we perform a maximum likelihood fit over the entire sky, placing  $\vec{\Omega}$ at the center of each bin of the Nside=64 HEALPix map. The maximization is done for $n_s$ only, and we obtain a best-fit value $\hat{n}_s$ for each $\vec{\Omega}$ and time window. For each GRB time window  we define the sky-map $\Lambda_{\nu}(\vec{\Omega})$:

\begin{eqnarray}
\Lambda_\nu(\vec{\Omega}) &=& 2 \cdot \log\left[\frac{\mathcal{L}(\hat{n}_s,\vec{\Omega})}{\mathcal{L}(n_s = 0, \vec{\Omega})} \right ] \nonumber \\
&=& -2 \hat{n}_s + 2 \sum_{i=1}^N \log\left [\frac{\hat{n}_s S(\mathbf{x}_i,\vec{\Omega})}{n_b B(\mathbf{x}_i,\vec{\Omega})}+1 \right ].
\end{eqnarray}

Note that the sky-map $\Lambda_\nu(\vec{\Omega})$ depends only on neutrino data. Up to this point, we have not used GRB positions or localization uncertainties. To include this information, and similarly to \citet{2022ApJ...939..116A}, we use the previously described \textit{Fermi-GBM} HEALPix maps, or synthetic maps, as a prior on the final test statistic. We define a weight  $w(\vec{\Omega}) = P(\vec{\Omega})/P_{max}$ for each GRB map. Here $P(\vec{\Omega})$ is the GRB localization probability evaluated at each sky location $\vec{\Omega}$, and $P_{max}$ is the maximum probability for the map of this GRB. The combined sky-map of neutrino and GRB information, $\Lambda_{final}(\vec{\Omega})$ is defined as:
\begin{eqnarray}
\Lambda_{final}(\vec{\Omega}) &=& 2 \cdot \log\left[w(\vec{\Omega}) \frac{\mathcal{L}(\hat{n}_s,\vec{\Omega})}{\mathcal{L_\nu}(n_s = 0,\vec{\Omega})} \right ] \nonumber
\\&=& \Lambda_{\nu}(\vec{\Omega}) + 2 \log(w(\vec{\Omega})). 
\end{eqnarray}

This all-sky map has a maximum value at $\hat{\vec{\Omega}}$, which for each GRB and each time window, defines the test statistic used in this work,
\begin{equation}
    TS=\Lambda_{final}(\hat{\vec{\Omega}}).
    \label{eq:ts}
\end{equation}

To estimate the significance of $TS$, we use a large number of scramblings. This is a standard technique in IceCube, in which the time of each event is randomized within the live time of IceCube being studied. This procedure keeps the reconstructed neutrino direction detector coordinates constant but randomizes the reconstructed neutrino right ascension. Given the very large duration of the live time compared to the total duration of all the time windows that we study, each scrambling dataset is a good approximation of the background expectation. In this work, we have used 1000,000 scramblings. 

The significance of correlations between neutrino events and each GRB for a given time window is quantified as a $p$-value by comparing the measured $TS$ to the background-only $TS$ distribution. The background-only $TS$ distribution is obtained via scramblings.

At this stage, we have a total of six $p$-values for each GRB, corresponding to the six time windows that have been studied. For each GRB we choose the most significant of the six $p$-values. We calculate a trial correction for the look-elsewhere effect for this choice by using the scrambled data cumulative distribution function of the $p$-value of the most significant time window. 
For each GRB we now have the most significant time window for correlation and the best fit $\hat{n}_s$ for this time window and a single per-GRB $p$-value.

Finally, the most significant per-GRB $p$-value is selected and trial-corrected to account for 2268 GRBs in the search. This results in the most significant GRB, its most significant time window, the best fit $\hat{n}_s$ for this GRB time window, and the post-trial $p$-value.

To characterize the performance of this analysis, we use sensitivity and $5\sigma$ discovery potential. With repeated simulated signal injected over background scramblings, we can calculate the signal $TS$ distribution. 
The sensitivity is the signal strength needed so that 90\% of the signal-injected $TS$ is above the median $TS$ for background scramblings. The 5$\sigma$ discovery potential is defined as the signal strength required for the injected median $TS$ to be greater than the $5\sigma$ threshold of the background scrambling $TS$ distribution. 

\begin{figure}[t]
\centering
\includegraphics[width=0.47\textwidth]{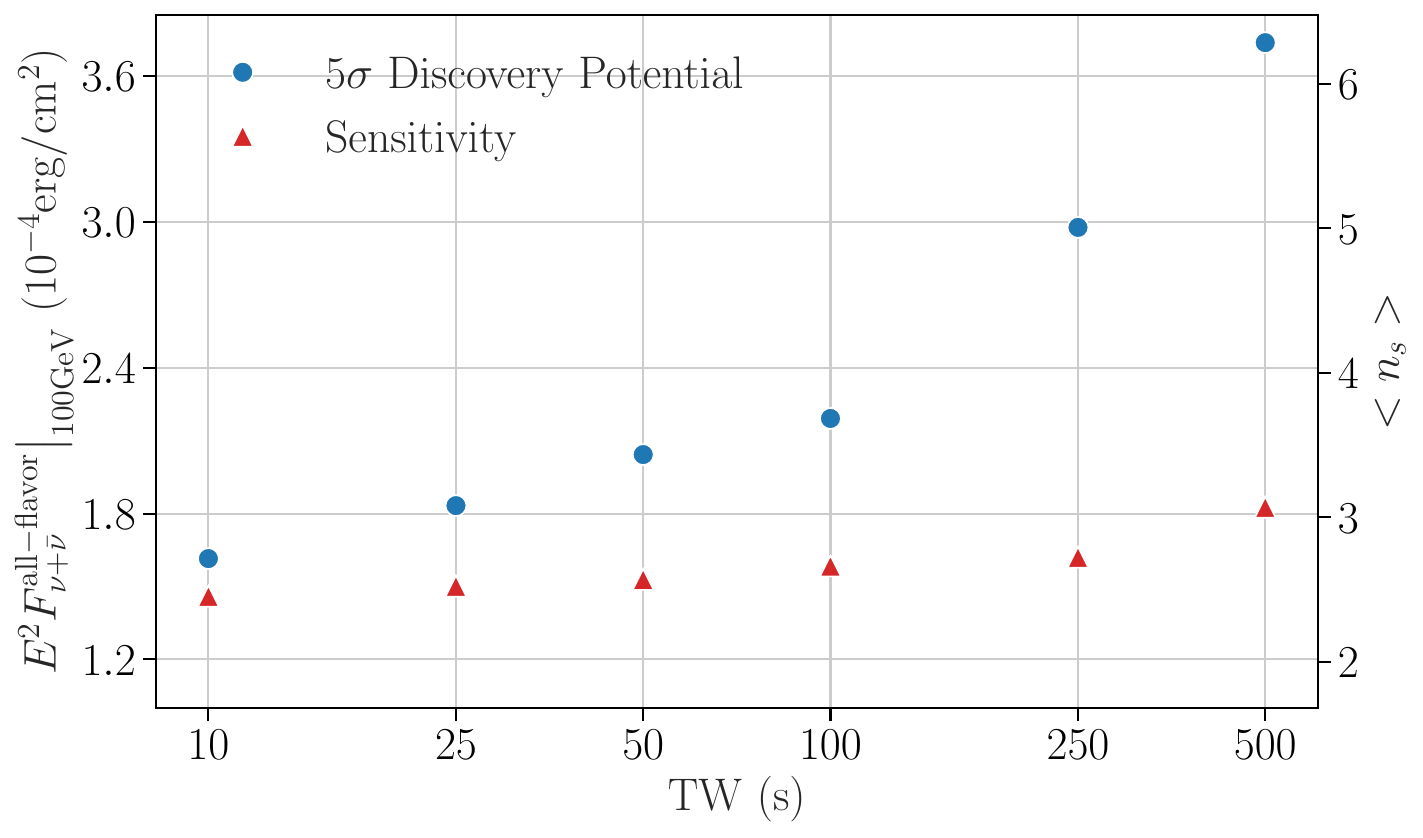}
\caption{Sensitivity and 5$\sigma$ discovery potential for (\textit{Fermi-GBM} detected) GRB bn~140807500 for the six time windows considered in this work.
This is the most significant GRB-neutrino correlation identified in this work.
The left ordinate axis is the all-flavor time-integrated neutrino number flux, $F_{\nu+\bar{\nu}}(E)$ times $E^2$ assuming an $E^{-2.5}$ spectrum and evaluated at 100~GeV. The right ordinate axis is the average number of signal events $<n_s>$ that result in the left ordinate axis. Here, TW is the duration of each time window. The sensitivity and 5$\sigma$ discovery potential for GRECO-Astronomy are only weakly dependent on declination or GRB localization uncertainty, therefore GRB bn~140807500 is representative of all GRBs studied here.\label{fig:sens_disc}}
\end{figure}

\begin{figure*}[t]
\centering
\includegraphics[width=0.32\textwidth]{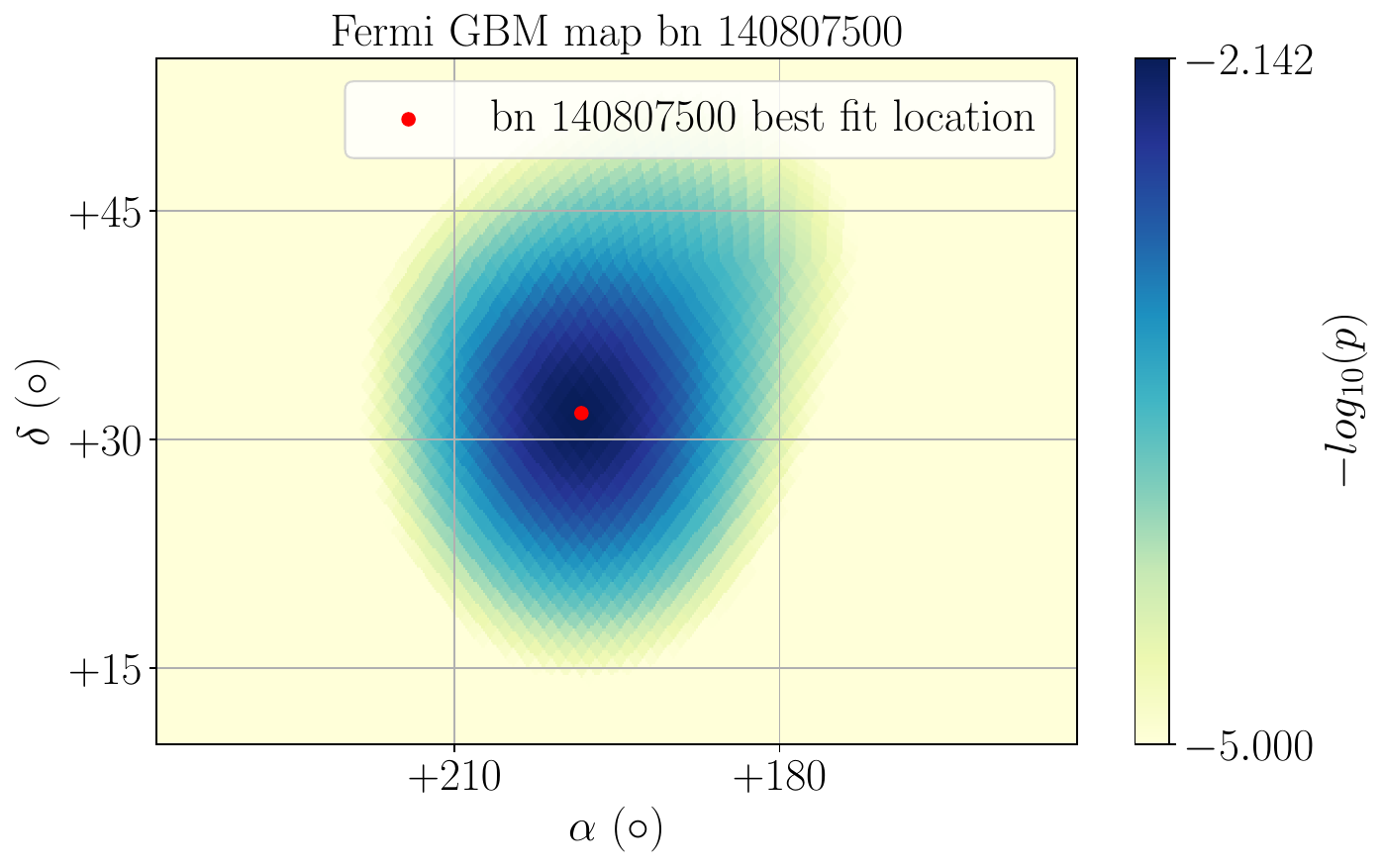}
\includegraphics[width=0.32\textwidth]{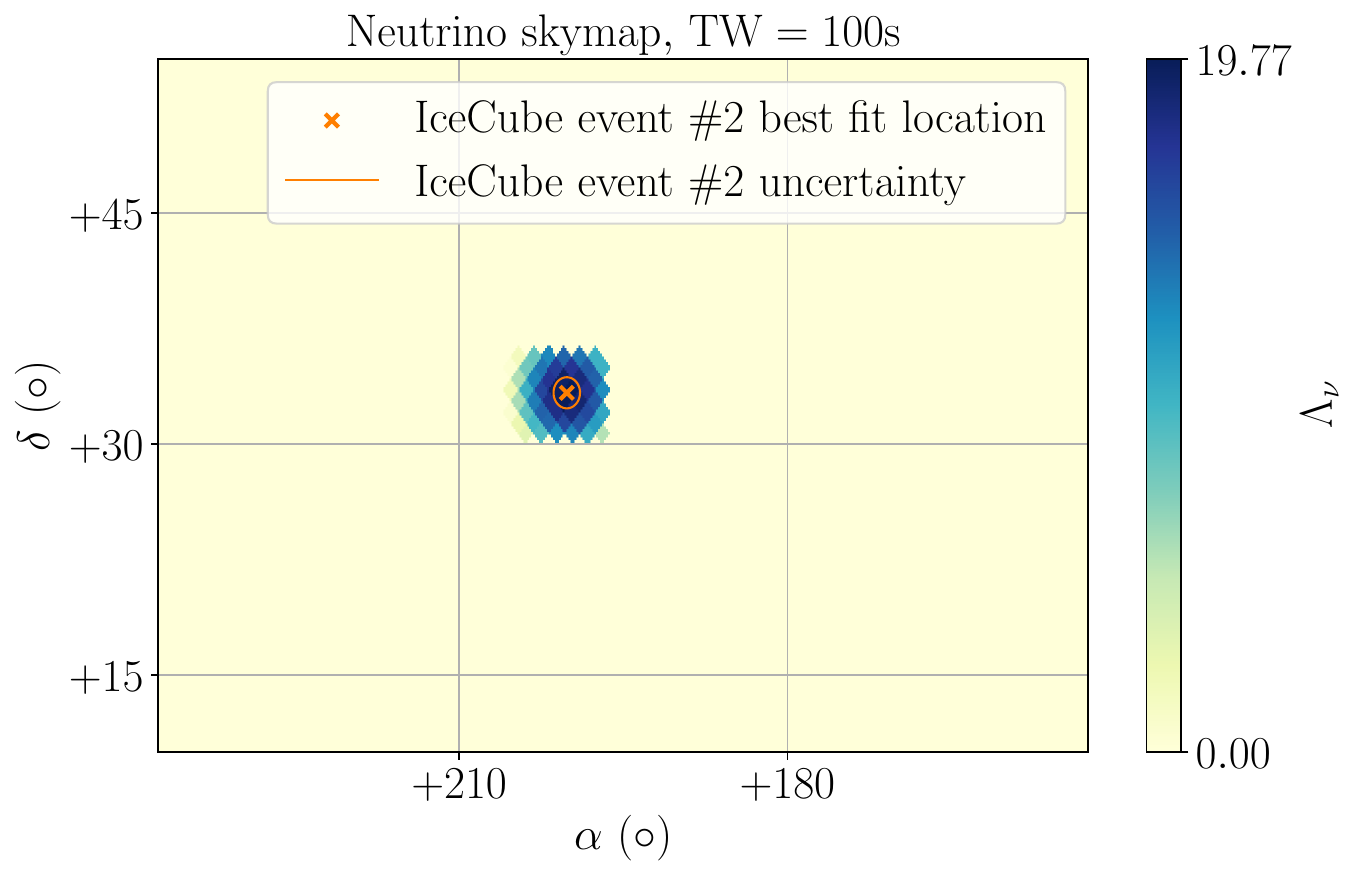}
\includegraphics[width=0.32\textwidth]{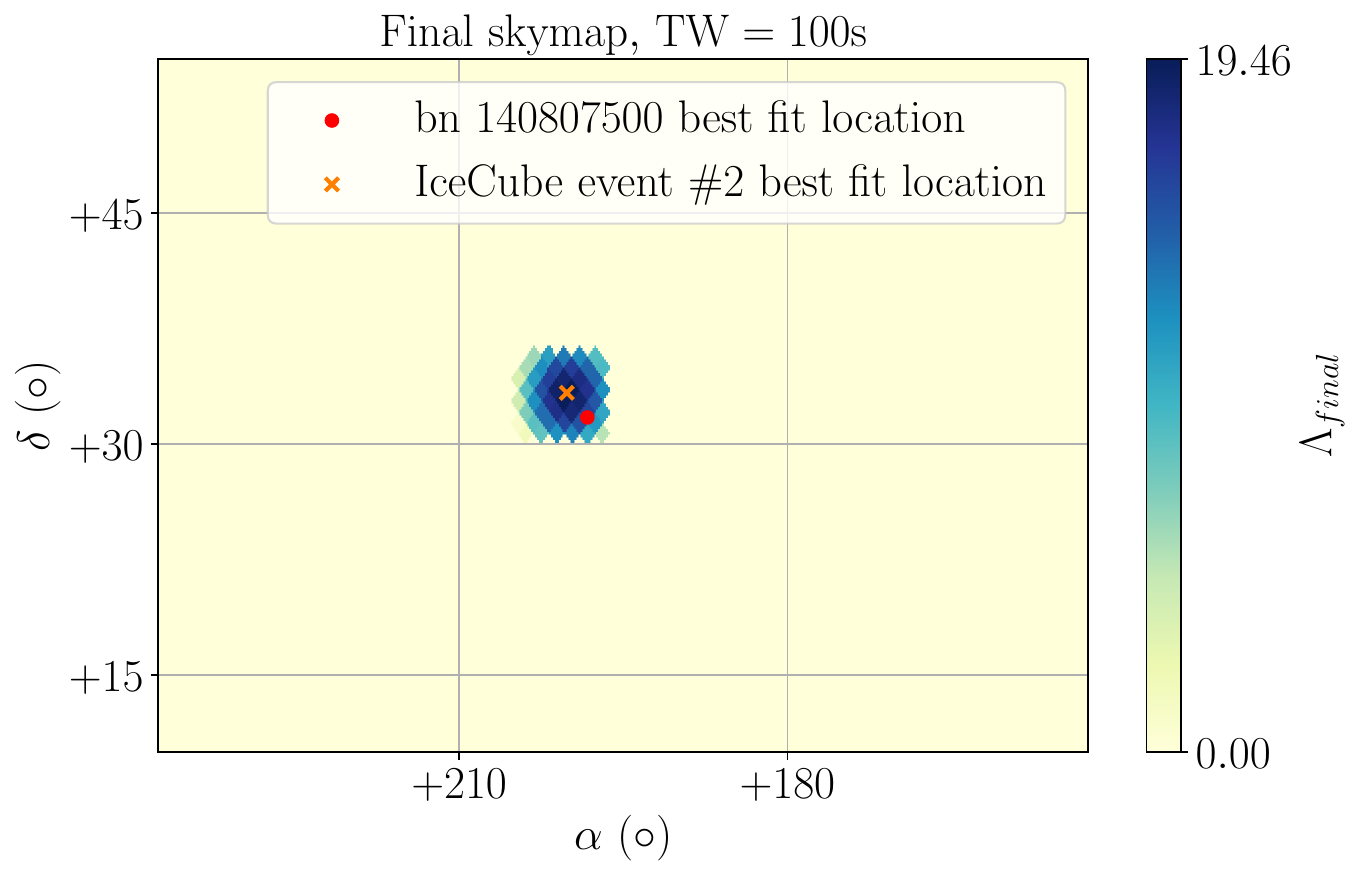}
\caption{Three skymaps in equatorial coordinates. (Left): The HEALPix skymap of the \textit{Fermi-GBM} GRB bn~140807500, the most significant GRB identified in this work. The best-fit location of bn~140807500 is that of $P_{max}$ on the GBM map. (Center): Neutrino skymap $\Lambda_{\nu}$. The direction and location uncertainty of neutrino event \#2 found in coincidence with bn~140807500 is shown. (Right): The neutrino and GRB final skymap $\Lambda_{final}$, which is  calculated from Eq. \ref{eq:ts}.
\label{figure:skymaps}}
\end{figure*}

Figure \ref{fig:sens_disc} shows the 90\% sensitivity and $5\sigma$ discovery potential for all six time windows for the most significant GRB identified in this analysis, bn~140807500. Neutrino acceptance and background rates vary slowly with declination in GRECO-Astronomy. This results in sensitivity and $5\sigma$ discovery potential that vary by, at most, a factor of $\sim$2 from the northern sky (best) to the southern sky (worst) \citep{2022arXiv221206810A}.
This is unlike GRB studies with TeV--PeV neutrinos in which sensitivity and $5\sigma$ discovery potential for northern declinations is a factor of $\sim$20 better than for southern declinations \citep{2022ApJ...939..116A}. 
At the South Pole, the intense down-going cosmic ray muon background worsens southern sky sensitivity for TeV--PeV neutrinos compared to the northern sky.
As the background rate in the work presented here is very low, sensitivity and $5\sigma$ discovery potential do not depend strongly on the GRB localization uncertainty. This is because the angular uncertainty of GRECO-Astronomy events is typically larger than GRB localization uncertainty. For example, a GRB localized to $15^\circ$ has $\sim$15\% higher background rate than one localized to $1^\circ$. GRB localization uncertainty has a smaller influence on sensitivity and $5\sigma$ discovery potential than declination.

\subsection{Binomial Test}

We statistically combine per-GRB $p$-values for the most significant time window, to search for a subgroup of GRBs that may be significant neutrino emitters. This case can be interesting when each individual GRB is not statistically significant by itself. The binomial probability, which has been used by IceCube in other works, e.g., \citet{aartsen2020time}, is given by:

\begin{equation}
P(k) = \sum_{m=k}^{N} \frac{N!}{(N-m)!m!} {p_{k}}^{m} (1-p_{k})^{N-m}.
\end{equation}
Here the pre-trial binomial $p$-value $P(k)$ denotes the probability of $k$ or more GRBs with $p$-values smaller than $p_k$ appearing among the background, where $p_k$ is the $k$th smallest pre-trial $p$-value in the final GRB list. Taking into account the number of independent GRBs (2268), we adjust the probability of observing a given result by chance with the cumulative density function (CDF) made out of the most significant binomial $p$-values obtained from 100 million null hypothesis binomial tests. The smallest pre-trial binomial $p$-value $P(k)$ obtained at a size of subgroup $k$ is corrected for trials and reported as post-trial binomial $p$-value $P_{binom}$.

\subsection{Systematic Uncertainties}
\label{sec:systematics}
We have estimated systematic uncertainties by varying parameters that affect signal efficiency or background rate. These include the optical properties of glacial ice, such as scattering and absorption; the relative DOM efficiency; the optical properties of the refrozen ice column in each IceCube string, aka ``hole ice'', resulting from the ice drilling; and the seasonal variations \citep{2023arXiv230304682A} on the rate of GRECO-Astronomy events. Following \citep{2022arXiv221206810A} we simulated changes of $\pm$10\% in DOM efficiency, $\pm$10\% in the absorption coefficient, $\pm$10 effective scattering coefficient, and $\pm$1$\sigma$ variations in hole ice optical properties. Because background is estimated via scrambling only seasonal variations are relevant as systematics. Signal, on the other hand, is affected by the other parameters described.  

To estimate systematic uncertainties for signal, we conducted simulations assuming a subphotospheric neutron-proton collision model with $\Gamma=300$ and a time search window of 2200~s. These values are appropriate for GRB~221009A. We find that the uncertainty in absorption length can degrade the sensitivity by 6.6\%; effective scattering length by 4.5\%, DOM efficiency by 3.8\%, and hole ice by 8.7\%. 

To evaluate the effect of seasonal variation, we considered the 1$\sigma$ background rate range for GRECO-Astronomy. Over the year, this corresponds to $\pm$10\% around the average of 4.6~mHz or 16.6 events per hour (before the final $E_{reco}>10$~GeV cut). We implemented a method in which we changed the local background rate by $\pm$10\% compared to the scrambled background rate. We used an ensemble of background scramblings with three different background rates for the pre-computed background scans. We find that seasonal variations can change sensitivity by 3\%.

We add in quadrature all the systematics uncertainties and find that the sensitivity can be degraded by 13\%. Performing a similar procedure for the $5\sigma$ discovery potential, we obtain that it can be degraded by 14\%. We adopt both these values to degrade limits, sensitivities, and 5$\sigma$ discovery potentials presented here. 

We have cross-checked these systematic uncertainty estimates with two other scenarios. In one scenario the Lorentz factor was increased to $\Gamma=800$. In this case, the total systematic uncertainty on sensitivity was reduced from 13\% to 11\%. We attribute the difference to the higher energy of events for a larger Lorentz boost factor. In the second scenario, we changed the time of the search from 2200~s to 221.1~s, the $T_{90}$ of GRB~221009A. In this case, the systematic for the sensitivity was reduced from 13\% to 10\%. A lower systematic is found for shorter time windows which is consistent with \citet{2022arXiv221206810A} which had larger systematic uncertainties but considered time windows of weeks-time scale. The $\Gamma=300$ and search window of 2200~s are conservative as they correspond to the lowest energy events and to the longest time window we consider. 

\section{Results} \label{sec:results}

\begin{deluxetable*}{ccccccccccccc}
\centerwidetable 
\tablewidth{\textwidth}
\tabletypesize{\footnotesize}
\tablenum{1}
\tablecaption{Top three most significant GRBs in the individual search analysis. The table shows the GRB name; start time of $T_{prompt}$; right ascension; declination; circularized localization uncertainty; most significant time window duration; best-fit $\hat{n}_s$ number of neutrino events in coincidence with the GRB; all-flavor neutrino number flux $F_{\nu+\bar{\nu}}(E)$ upper limit times $E^2$ for the most significant time window $TW$ for two spectral indices; whether or not a synthetic prior was used; the GRB pre-trial $p$-value, corrected for the 6-time windows used in this analysis; and the post-trail $p$-value for the most significant burst.  The right ascension and declination for bn~140708500 and bn~160804968 are for $P_{max}$ of the GBM map.
\label{tab:GRBlist}}
\tablewidth{0pt}
\tablehead{
\colhead{GRB} & \colhead{Start time}  & \colhead{RA} & \colhead{Dec} & \colhead{Loc.} & \colhead{Loc.} & \colhead{TW} & \colhead{$\hat{n}_s$} &  \multicolumn{2}{c}{$E^2 F_{\nu+\bar{\nu}}(E)$ Upper Limit} & \colhead{Synth.} & \colhead{Pre-trial} & \colhead{Post-trial}\\
\colhead{Name} & \colhead{} & \colhead{} & \colhead{} & \colhead{Unc.} & \colhead{Satellite} & \colhead{} & \colhead{} &  \colhead{$\gamma=2.5|_{100\mathrm{GeV}}$} & \colhead{$\gamma=2.0$} &\colhead{prior} & \colhead{$p$-value} & \colhead{$p$-value}
\\
\colhead{} & \colhead{(UTC)} & \colhead{($^\circ$)} & \colhead{($^\circ$)} & \colhead{($^\circ$)} & \colhead{} & \colhead{(s)} & \colhead{} & \multicolumn{2} {c}{(10$^{-5}$erg\,cm$^{-2}$ s$^{-1}$)} &\colhead{} & \colhead{} & \colhead{}
}
\startdata
bn~140807500 & \begin{tabular}[c]{@{}l@{}}2014-08-07\\ 11:59:33\end{tabular}  & 198.3 & 31.7 & 4.4 &\textit{GBM} & 100 & 1.08  & 12.84 & 5.24 & No & $4.6\times10^{-6}$ & 0.097\\
bn~160804968 & \begin{tabular}[c]{@{}l@{}}2016-08-04\\ 23:13:34\end{tabular}  & 82.3 & -23.0 & 9.1 & \textit{GBM} & 100 & 2.81  &  23.04 & 11.94 & No & $9.6\times10^{-4}$ & --\\
GRB~160802A & \begin{tabular}[c]{@{}l@{}}2016-08-02\\ 06:13:30\end{tabular}  & 28.1 & 71.4 & 1.0 & \textit{IPN} & 50 & 2.74  &  16.92 & 8.26 & Yes  & $1.3\times10^{-3}$ & --
\enddata
\end{deluxetable*}

The most significant GRB-neutrino correlation identified among 2268 GRBs is the \textit{Fermi-GBM} GRB bn~140807500. The $p$-value, corrected for the 6 time windows, is $4.6 \times 10^{-5}$ and is found for a time window of 100~s with the best-fit number of events $\hat{n}_s = 1.08$. After correcting for trials for 2268 GRBs we obtain a post-trial $p$-value of $0.097$ ($1.3\sigma$). Burst bn~140807500 triggered \textit{Fermi-GBM} on August 7, 2014. This is a short GRB with $T_{90}=0.512\pm0.202$~s and an energy fluence of ($1.289\pm0.014)\times10^{-6}$~erg/cm$^{2}$ \citep{grbweb}. Because only \textit{Fermi-GBM} identified this burst, the $T_{90}$ matches $T_{prompt}$. 

This analysis uses the \textit{Fermi-GBM} maps whenever they are available, where the most likely location of this GRB is right ascension $\alpha=$198.3$^\circ$ and declination $\delta=31.7^\circ$.
Two neutrino candidates are identified in IceCube-DeepCore in the 100~second time window around the center of $T_{prompt}$ for bn~140807500. Event one is identified as a cascade detected 33.67 seconds before the center of the $T_{prompt}$ window. This cascade has a reconstructed energy of 50~GeV. The best-fit reconstructed direction is $\alpha=76.8^\circ$ and $\delta=-51.5^\circ$, which is $133.5^\circ$ away from the most likely GRB location. Event one has a circularized directional uncertainty of $35.4^\circ$. Event two is a starting track detected 38.58 seconds after the center of the $T_{prompt}$ window. The reconstructed energy is 221~GeV. The best-fit reconstructed direction is $\alpha=200.2^\circ$ and $\delta=33.3^\circ$, which is $2.3^\circ$ away from the most likely GRB location. Event two has a circularized directional uncertainty of  $1.0^\circ$. Figure \ref{figure:skymaps} shows the localization probability skymap of bn~140807500 provided by \textit{Fermi-GBM}, the $\Lambda_\nu$ all-sky scan from neutrino events within the 100~s time window, and the $\Lambda_{final}$ skymap. Figure \ref{figure:skymaps} shows that the second neutrino candidate is responsible for a relatively high test statistic. Table \ref{tab:GRBlist} shows the three most significant GRB-neutrino coincidences identified. 

The smallest binomial $p$-value is $P(k=1)=0.099$ with index $k=1$ with threshold $p$-value $p_k = 4.6 \times 10^{-5}$. After correcting for trials due to testing multiple thresholds, the post-trial binomial $p$-value of 0.65 is obtained. So no additional neutrino event-GRB correlations, besides bn~140807500, are identified with the binomial test. 

\section{Constraint on baryon loading} \label{sec:baryon loading}

We have followed calculations by \citet{2013PhRvL.111m1102M} and \citet{2022ApJ...941L..10M}, for the neutron-proton collision scenario, to obtain an all-flavor neutrino signal expectation for the GRBs used in this work. The authors of these works provided us with neutrino spectra which allow the calculation of $E^2 F(E)$ \citep{MuraseGitHub}. Given a GRB, the model depends on the gamma-ray energy fluence of the prompt phase $f_\gamma$, redshift $z$, and the product of baryon loading times the neutron-proton opacity $\xi_N \tau_{np}$. Following \citet{2013PhRvL.111m1102M} we adopt the definition of the baryon loading as the ratio of the equivalent isotropic energies in baryons to gamma rays,  $\xi_N = E_{N,iso}/E_{\gamma,iso}$, with $\xi_N=5$ and $\tau_{np}=1$ as default model values.

Of the 2268 GRBs studied here, four do not have a measured energy fluence in GRBWeb so we exclude these GRBs from the calculation. Only 198 GRBs have a measured redshift, of which 187 are long and 11 are short GRBs. Among these, the mean redshift for long GRBs is 2.0 and, for short GRBs is 0.7. In our calculations for GRBs without a measured redshift we adopt a default value of $z=0.7$ for short GRBs and $z=2.0$ for long GRBs. With a Lorentz factor of $\Gamma=300$ (800), and default $\xi_N \tau_{np}=5$ we obtain an all-flavor neutrino signal, using GRECO-Astronomy simulation with $E_{reco}>$10~GeV, in IceCube-DeepCore of 0.64 (1.39) events for all 2264 GRBs combined.  As a systematic check, we have redone the calculation for 2264 GRBs by assuming $z=0.25$ and $z=1.0$ for short and long GRBs respectively, that do not have a measured redshift. We find for this check a signal expectation of 0.64 (1.24). Therefore we find that the lack of redshift information for most GRBs does not significantly affect the neutrino signal prediction for the sub-photospheric model. 

The all-flavor signal expectation for 2264 GRBs can be compared to that for GRB~221009A. For the latter, we use z=0.151 \citep{2022GCN.32648....1D}, an energy fluence measured by \textit{Konus-Wind} of $f_\gamma = 1.2\times10^{55}$~erg \citep{2023ApJ...949L...7F}, $\xi_N \tau_{np}=5$ and Lorentz factor of 300 (800). This results in an expectation of 4.72 (8.56) events. Alternatively, using the energy fluence measured by \textit{Fermi-GBM}, $f_\gamma = 1.0\times10^{55}$~erg \citep{2023arXiv230314172L}, results in a factor $\times1.2$ lower signal expectation, as $F(E)$ is proportional to $f_\gamma$.

\begin{figure}[t]
\centering
\includegraphics[width=0.47\textwidth]{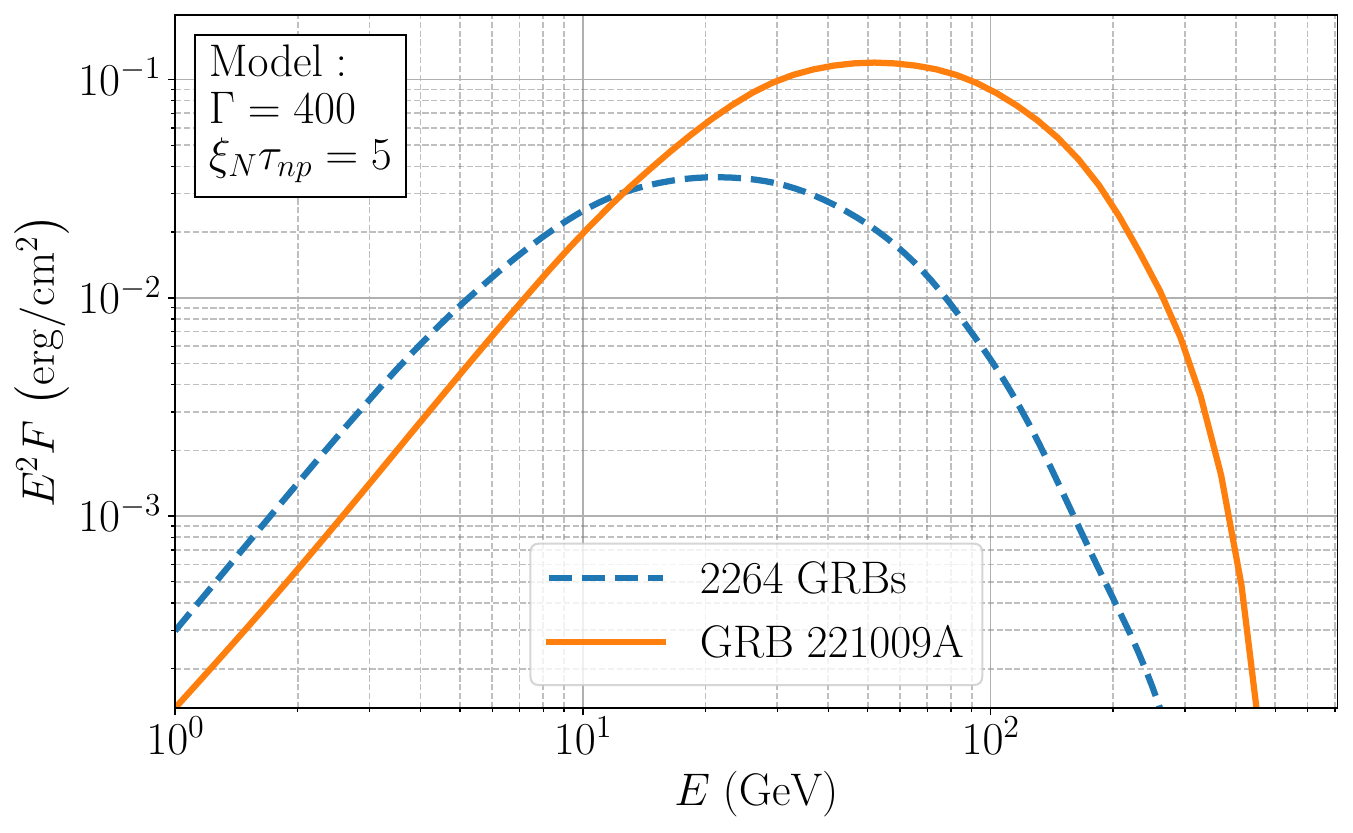}
\caption{Comparison of all-flavor neutrino time-integrated neutrino number flux, $F(E)$ times $E^2$ for GRB~221009A and 2264 GRBs with measured energy fluence. The calculation follows \citet{2013PhRvL.111m1102M} and \citet{2022ApJ...941L..10M} and assumes a Lorentz  factor, $\Gamma=400$ and $\xi_N \tau_{np} = 5.$
 \label{fig:GRBcomp}}
\end{figure}

\begin{figure*}[t]
\centering
\includegraphics[width=0.47\textwidth]{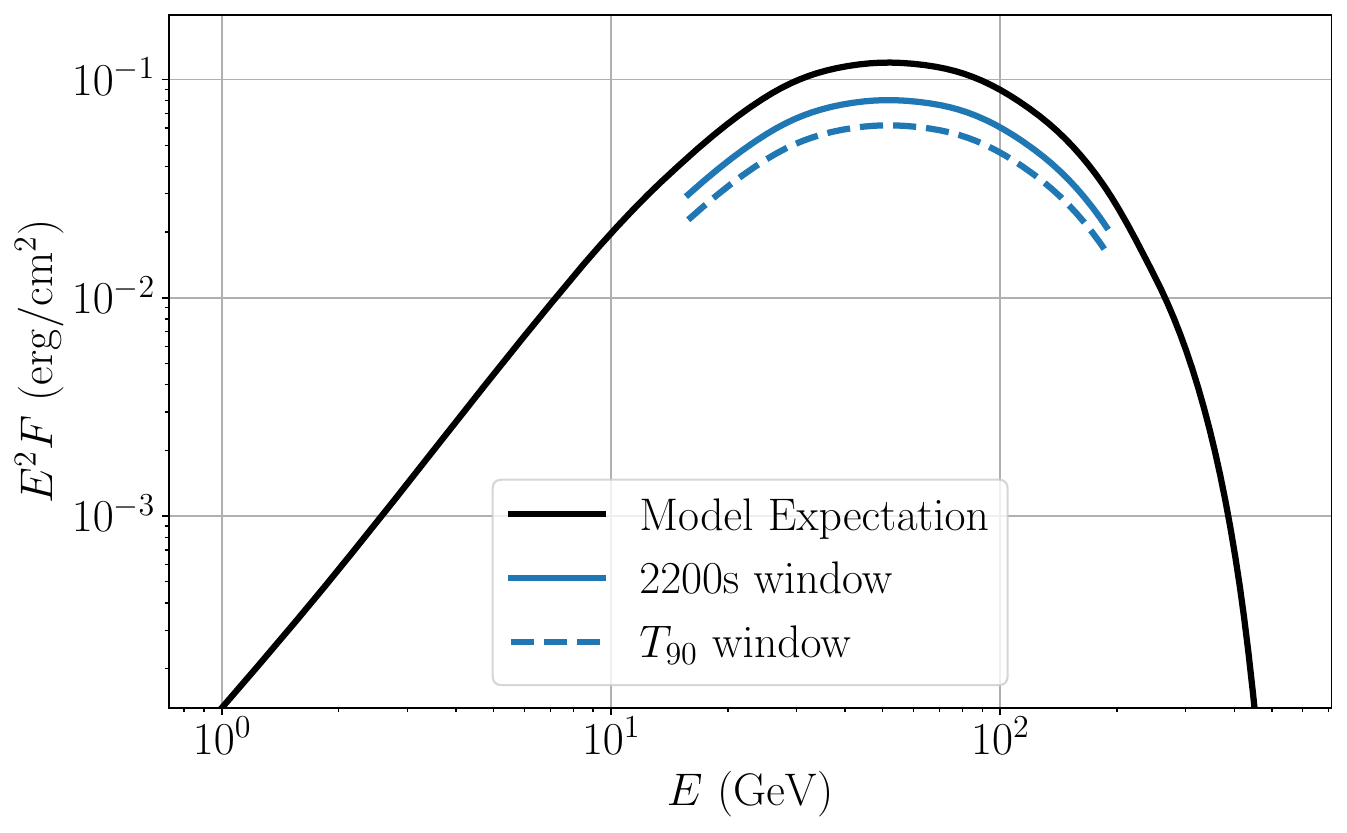}
\includegraphics[width=0.48\textwidth]{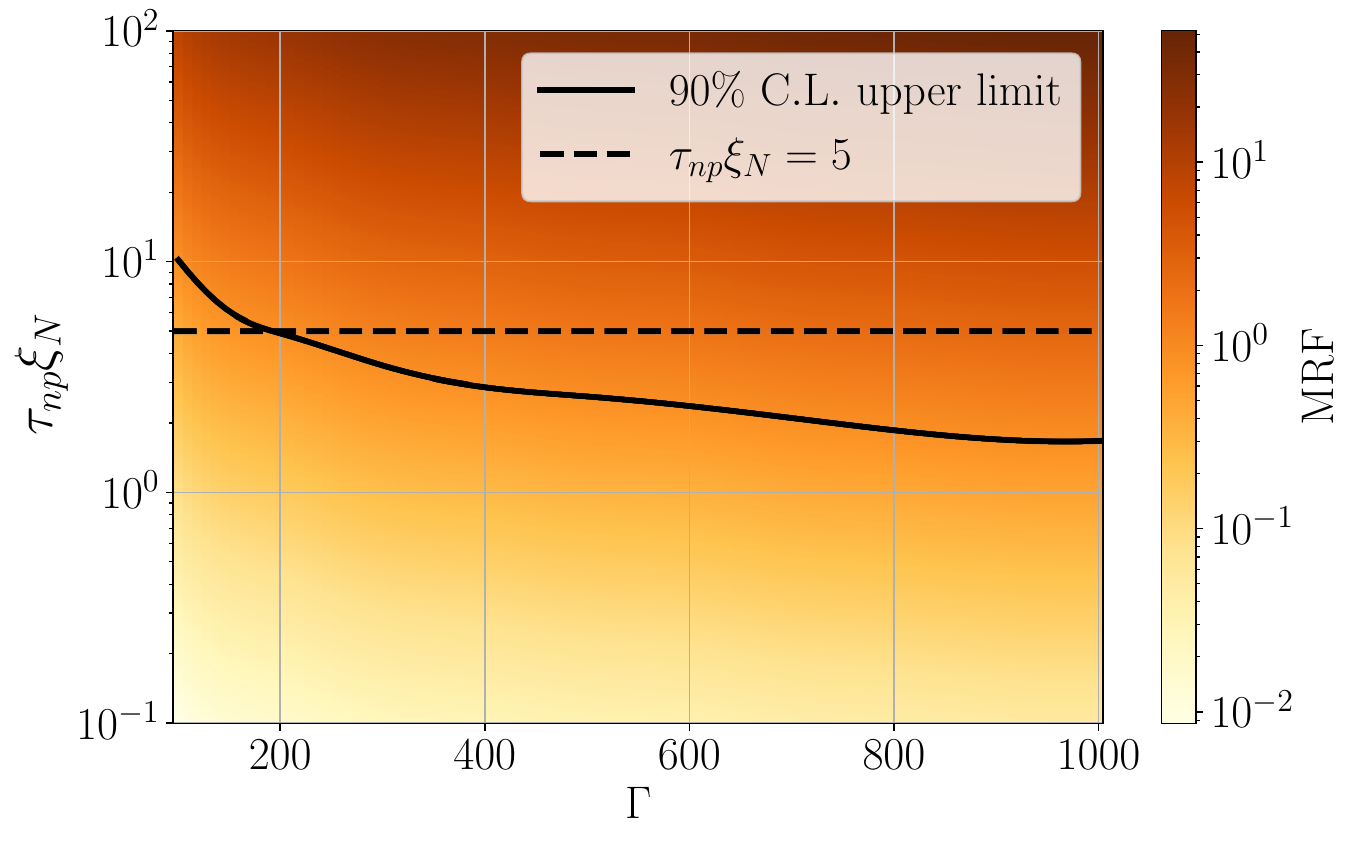}
\caption{Sub-photospheric neutron-proton collision model and limits for GRB~221009A. (Left) All-flavor sub-photospheric model prediction for GRB~221009A assuming a Lorentz factor of $\Gamma=400$, an equivalent isotropic energy of $1.2\times10^{55}$~erg and neutron-proton opacity times baryon loading $\xi_N\tau_{np}=5$. GRECO-Astronomy 90\% confidence level limits are during $T_{90}$ and during a 2200~s time window that includes 200~s prior to the start of $T_{90}$  \citep{icrc2023:grb221009a}. (Right) Constraint on $\xi_N\tau_{np}$, as a function of the bulk jet Lorentz factor $\Gamma$ for GRB~221009A. The model rejection factor (MRF) is the ratio of model signal event expectation to the 90\% confidence level limit on the signal. We use the GRB~221009A limit from a time window of 2200~s. The region above the black solid line is rejected at a 90\% confidence level. The canonical value of $\xi_N\tau_{np}=5$ is shown as a dashed black line.
\label{fig:XivsGamma}}
\end{figure*}

Remarkably, GRB~221009A has a larger neutrino signal expectation in IceCube-DeepCore than the other 2264 GRBs combined, by a factor of 6 to 8. Even if we assume redshift values for bursts without a redshift measurement that are significantly smaller than typical, we still find that GRB~221009A has a larger signal expectation over the other 2264 GRBs combined by a similar factor. A comparison of $E^2 F(E)$ for GRB~221009A and 2264 GRBs combined is shown in Figure \ref{fig:GRBcomp}. There are two reasons why GRB~221009A has a larger neutrino signal expectation than the combination of the GRBs studied here. First, neutrinos of higher energy are easier to identify and correlate with GRBs because they typically have lower angular uncertainty and because the detector has a higher efficiency for detecting them. 
The peak energy of $E^2 F(E)$, seen in Figure \ref{fig:GRBcomp}, is inversely proportional to $(1+z)$ and the redshift of GRB~221009A is small. So GRB 221009A has higher energy neutrinos than a GRB at $z=2$. Second, GRB~221009A has an extremely high energy fluence, and the height of the peak of $E^2 F(E)$ seen in Figure \ref{fig:GRBcomp} is proportional to $f_\gamma$.

The parameter product $\xi_N\tau_{np}$ is unknown for all GRBs. We assume  $\tau_{np}=1$, which leaves the baryon loading unknown. Knowledge of the baryon loading can provide information about the environment for the formation of the jet. The best constraint on $\xi_N\tau_{np}$ can be derived from GRB~221009A for two reasons: GRB~221009A's high signal expectation compared to the set of 2264 GRBs and the lack of correlated neutrino events reported in \citet{2023ApJ...946L..26A}; and because for 2264 GRBs, with a per GRB window of 500 s, background accumulates over 1,132,000 s (13 days), while for GRB 221009A the background is only over a single time window of 2200 s. For GRB~221009A we calculate the constraint on $\xi_N$ using the 2200~s time window, as this includes observations prior to the prompt phase. 
Stating that the constraint from GRB~221009A is the most restrictive, assumes that GRB~221009A has similar characteristics as all other GRBs. This assumption may not be valid, e.g., because of the unusually high isotropic equivalent energy of GRB~221009A. Therefore there is still important, complementary information provided by the neutrino flux constraints we have placed on the other 2268 GRBs.

With a Lorentz factor of 300 (800) and a time window of 2200~s, the upper limit on $E^2 F(E)$ for GRB~221009A is a factor of 1.65 (2.98), lower than predicted by the model in \citet{2022ApJ...941L..10M}, but updated to use the \textit{Konus-Wind} energy fluence. This upper limit directly translates into a constraint of $\xi_N\tau_{np}<$~3.85 (2.13) for a Lorentz factor of 300 (800). In Figure \ref{fig:XivsGamma} the left panel shows the canonical model prediction for GRB~221009A and GRECO-Astronomy limits from \citet{icrc2023:grb221009a} for a Lorentz factor of $\Gamma=400$. The right panel shows the constraint on $\xi_N \tau_{np}$ as a function of the Lorentz factor.

\citet{2023arXiv230314172L} provide lower bounds on the Lorentz factor of GRB~221009A under various modeling scenarios. Requiring no photon-photon $e^+e^-$ pair production for the highest energy photons observed in \textit{Fermi-GBM} and using a single zone model, they set a constraint of $\Gamma \gtrsim 1560$. However, under more realistic scenarios, the lower bound is a factor of 2 lower or $\Gamma \gtrsim 780$. Figure \ref{fig:XivsGamma} is agnostic to the choice of Lorentz factor, but taking a value of $\Gamma=780$, results in a constraint on the baryon loading that is significantly lower than the canonical theoretical value. 

The constraint on baryon loading we present here is complementary to that which can be set, also for GRB~221009A, using TeV--PeV neutrinos. With 10--1000 GeV neutrinos, the limit on $\xi_N\tau_{np}$ is more constraining as the Lorentz factor increases. 
This is because the peak value of $E^2 F(E)$ does not depend on the Lorentz factor but the energy for which $E^2 F(E)$ peaks depends linearly on the Lorentz factor. A higher value of the Lorentz factor results, on average, on higher-energy neutrinos, which are easier to detect and correlate to a GRB. 
The situation is reversed for TeV--PeV neutrinos in which the constraint on the baryon loading is best for low values of the Lorentz factor.
For both the internal shock model and the Internal-Collision-Induced Magnetic Reconnection and Turbulence, ICMART, \citep{2011ApJ...726...90Z,2013PhRvL.110l1101Z} model which predict TeV--PeV neutrinos, the peak value of $E^2 F(E)$ is lower for larger values of the Lorentz factor.
Using TeV--PeV neutrinos, the 90\% confidence level upper limit on the baryon loading, derived from GRB 221009A, under the internal shock model and for $\Gamma=300$, is $\xi_N<0.55$. Under the same assumptions, but for the ICMART  model, the limit is  $\xi_N<2.97$ \citep{icrc2023:grb221009a}. 

\begin{figure}[t]
\centering
\includegraphics[width=0.47\textwidth]{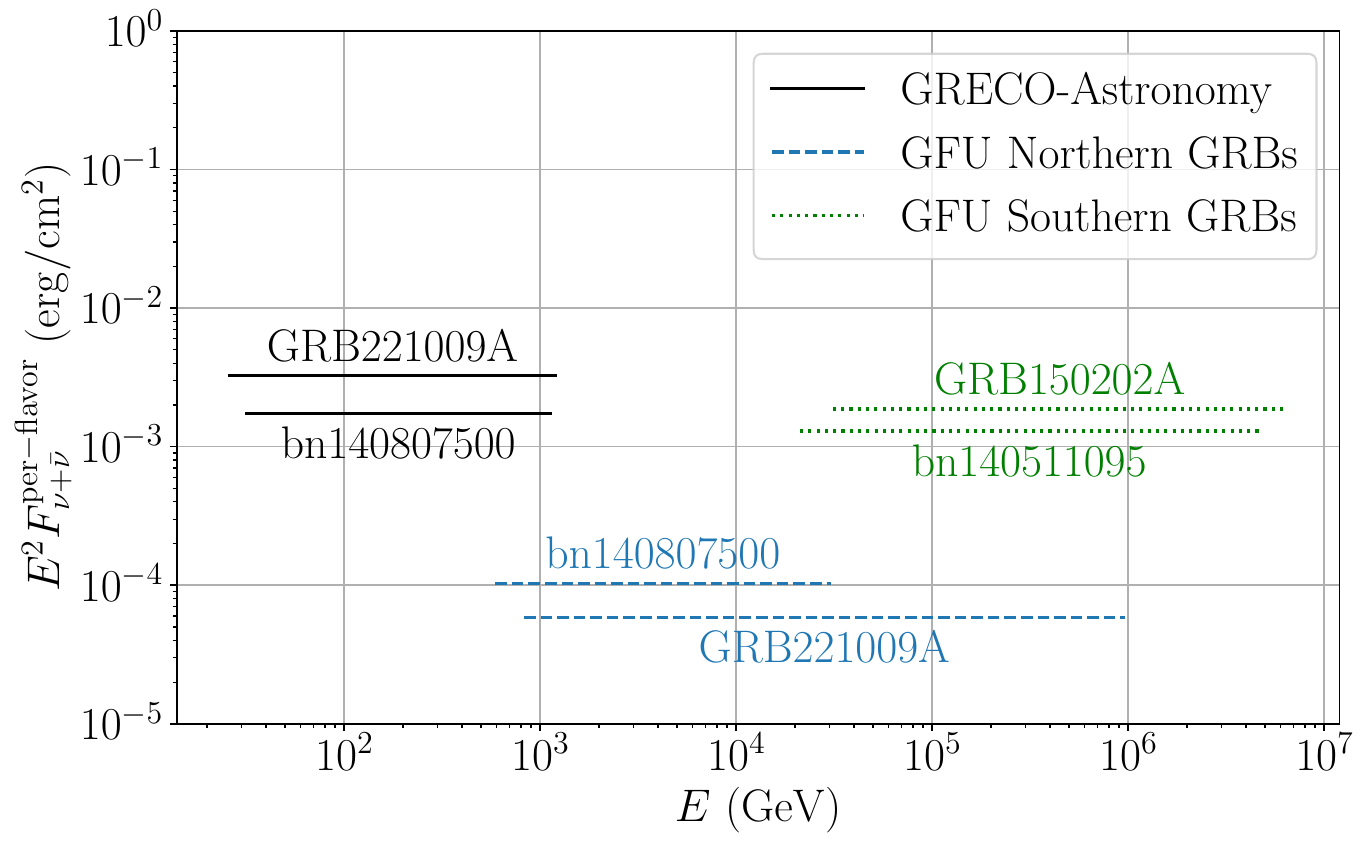}
\caption{Limits on per-flavor time-integrated neutrino number flux, $F_{\nu+\bar{\nu}}(E)$, times $E^2$ for six GRBs. 
Black lines show two GRBs studied with GRECO-Astronomy. One GRB is bn 140807500 (TW=100~s), the most significant GRB identified in this work, and the other is GRB~221009A ($TW=T_{90,GBM}$) \citep{2023ApJ...946L..26A}. Blue lines are for two northern sky GRBs, studied with TeV--PeV, using the GFU (Gamma-ray Follow-Up) dataset. These are, again, bn~140807500 (for $TW$=100~s) \citep{2022ApJ...939..116A} and GRB~221009A ($TW=T_{90,GBM}$) \citep{2023ApJ...946L..26A}. Green lines are for two southern sky bursts: GRB~150202A ($TW$=2~days) and Fermi GRB bn~140511095 ($TW$=2~days) \citep{2022ApJ...939..116A}.
The energy ranges shown correspond to the central 90\% of neutrino energies that would contribute to an $E^{-2}$ signal. The sensitivity of GRECO-Astronomy varies by a factor of $\sim$2 as a function of declination. On the other hand, the sensitivity of TeV--PeV studies changes significantly from the northern hemisphere to the southern hemisphere. To derive the per-flavor GRECO-Astronomy limits a 1:1:1 flavor flux ratio has been assumed.
\label{fig:LizComparison} }
\end{figure}

\section{Comparison of results of prior IceCube GRB studies}

Figure \ref{fig:LizComparison} shows a comparison of limits on $E^2 F(E)$ for several GRBs set by IceCube using GRECO-Astronomy and TeV--PeV searches. 
It is worth noticing that the recent study of correlations and GRBs for $\gtrsim$TeV neutrinos \citep{2022ApJ...939..116A} also found a correlation between bn~140807500 and event two (See section \ref{sec:results}). In that work, bn~140807500 was the most likely GRB-neutrino correlation among short GRBs in the northern sky (which IceCube defines as $\delta>-5^\circ$). The dataset used on \citet{2022ApJ...939..116A} was collected from April 2012 to October 2018. During this data period, 2.9\% of the events in the data used in the current work are also found in \citet{2022ApJ...939..116A}. Because event two is an event that starts in DeepCore and has relatively high energy, it is not surprising that both analyses identify it.

\section{Conclusions} \label{sec:conclusions}

Using IceCube data and public GRB data, we have studied 10--1000~GeV neutrino and GRB correlations for 2268 GRBs detected over 8 years. No evidence for neutrino emission by GRBs is found using either of the two analysis methods. In the first method, we search for the most statistically significant GRB-neutrino correlation. The most significant GRB is \textit{Fermi-GBM} bn 140807500 with a post-trial p-value of 0.097. In the second method, we statistically combined the results for all 2268 GRBs to search for a set of GRBs that could be significant as a group but not individually. We do not find any additional burst, besides bn 140807500, to possibly contribute, and the p-value for this test is 0.65. 

We compare sub-photospheric model predictions for a subset of 2264 GRBs, with prompt gamma-ray energy-fluence measurements, to the sub-photospheric model prediction of the \textit{Brightest of all time} GRB 221009A. We find that in the subphotospheric model, GRB~221009A results in a neutrino signal in IceCube-DeepCore that is $\gtrsim$6 larger than for the combined set of 2264 GRBs. We use previously calculated limits on neutrino emission for GRB~221009A to constrain the baryon loading of the jet. For a Lorentz factor of 300 (800), the baryon loading on GRB 221009A is lower than 3.85 (2.13) at a 90\% confidence level. The set of 2268 GRBs may still be useful to constrain models besides the sub-photospheric model. 

While GRECO-Astronomy cannot reach the sensitivity that the TeV--PeV searches achieve for northern sky GRBs, it covers a complementary energy range where different physical mechanisms of neutrino emission can be explored.

Future work that benefits from the use of the IceCube-Upgrade \citep{icrc2021:Upgrade} will enhance the sensitivity of IceCube-DeepCore to 10--1000~GeV neutrinos as the angular resolution of reconstructed events is expected to improve.  

\begin{acknowledgments}

We are grateful to K. Murase for providing templates of neutrino spectra necessary to calculate neutrino fluxes for the sub-photospheric model under the collision scenario.
We are grateful to J. Wood, and A. Goldstein for suggestions that improved this manuscript. The authors gratefully acknowledge the support from the following agencies and institutions: USA – U.S. National Science Foundation-Office of Polar Programs, U.S. National Science Foundation-Physics Division, U.S. National Science Foundation-EPSCoR, U.S. National Science Foundation-Office of Advanced Cyberinfrastructure, Wisconsin Alumni Research Foundation, Center for High Throughput Computing (CHTC) at the University of Wisconsin–Madison, Open Science Grid (OSG), Partnership to Advance Throughput Computing (PATh), Advanced Cyberinfrastructure Coordination Ecosystem: Services \& Support (ACCESS), Frontera computing project at the Texas Advanced Computing Center, U.S. Department of Energy-National Energy Research Scientific Computing Center, Particle astrophysics research computing center at the University of Maryland, Institute for Cyber-Enabled Research at Michigan State University, Astroparticle physics computational facility at Marquette University, NVIDIA Corporation, and Google Cloud Platform; Belgium – Funds for Scientific Research (FRS-FNRS and FWO), FWO Odysseus and Big Science programmes, and Belgian Federal Science Policy Office (Belspo); Germany – Bundesministerium für Bildung und Forschung (BMBF), Deutsche Forschungsgemeinschaft (DFG), Helmholtz Alliance for Astroparticle Physics (HAP), Initiative and Networking Fund of the Helmholtz Association, Deutsches Elektronen Synchrotron (DESY), and High Performance Computing cluster of the RWTH Aachen; Sweden – Swedish Research Council, Swedish Polar Research Secretariat, Swedish National Infrastructure for Computing (SNIC), and Knut and Alice Wallenberg Foundation; European Union – EGI Advanced Computing for research; Australia – Australian Research Council; Canada – Natural Sciences and Engineering Research Council of Canada, Calcul Québec, Compute Ontario, Canada Foundation for Innovation, WestGrid, and Digital Research Alliance of Canada; Denmark – Villum Fonden, Carlsberg Foundation, and European Commission; New Zealand – Marsden Fund; Japan – Japan Society for Promotion of Science (JSPS) and Institute for Global Prominent Research (IGPR) of Chiba University; Korea – National Research Foundation of Korea (NRF); Switzerland – Swiss National Science Foundation (SNSF).
\end{acknowledgments}

\bibliography{bibliography}{}
\bibliographystyle{aasjournal}
\end{document}